\documentclass[article]{aastex631}

\def\beq{\begin{equation}}
\def\eeq{\end{equation}}
\def\I{{\sc i}}
\def\II{{\sc ii}}

\begin{document}


\title{The relation of metal-poor stars to nearby solar analogues}


\author{Charles R. Cowley}
\affiliation{Department of Astronomy, University of Michigan, 
1085 S. University, Ann Arbor, MI 481090-1107\\
orcid I.D.:0000-0001-9837-3662}                            

\author{Robert E. Stencel}
\affiliation{Chamberlin Observatory, University of Denver, 2930 E Warren Ave., 
     Denver, CO 80210, USA, \\
     orcid I.D.:0000-0001-8217-9435\\
     e-mail: robert.stencel@du.edu}

\begin{abstract}
Sunlike dwarf stars in the solar neighborhood reflect ages, an ``average'' 
chemical evolution, and departures from that average. We show the chemical, 
and kinematic properties of four groups of Sunlike dwarfs form a continuum 
related to age. We plot $[Fe/H]$ vs. age, as well as kinematical values for 
the four groups. The vertical (negative) scatter in $[Fe/H]$ increases with 
age in a systematic way: as the age increases, $[Fe/H]$ decreases. The sets of 
Solar and metal-poor stars in the solar neighborhood are related by 
distributions in $[Fe/H]$ vs. age, as well as in Galactic position (XYZ) and 
velocity space (UVW). Among the samples there are no clusters of points that 
set one sample apart from the others. The distributions vary slowly from one 
set to the next, suggesting a mixture of stellar populations. A plot in Energy 
vs angular momentum phase space, with coordinate origin moved to the 
Galactic center, highlights different aspects of the kinematics of the four 
groups of stars. We finally compare the kinematic properties of these
four groups with those of two sets of ultra metal-poor stars.
\end{abstract}

\section{Introduction\label{sec:intro}}
 \begin{figure}
 \includegraphics[width=14cm,height=10cm,keepaspectratio]{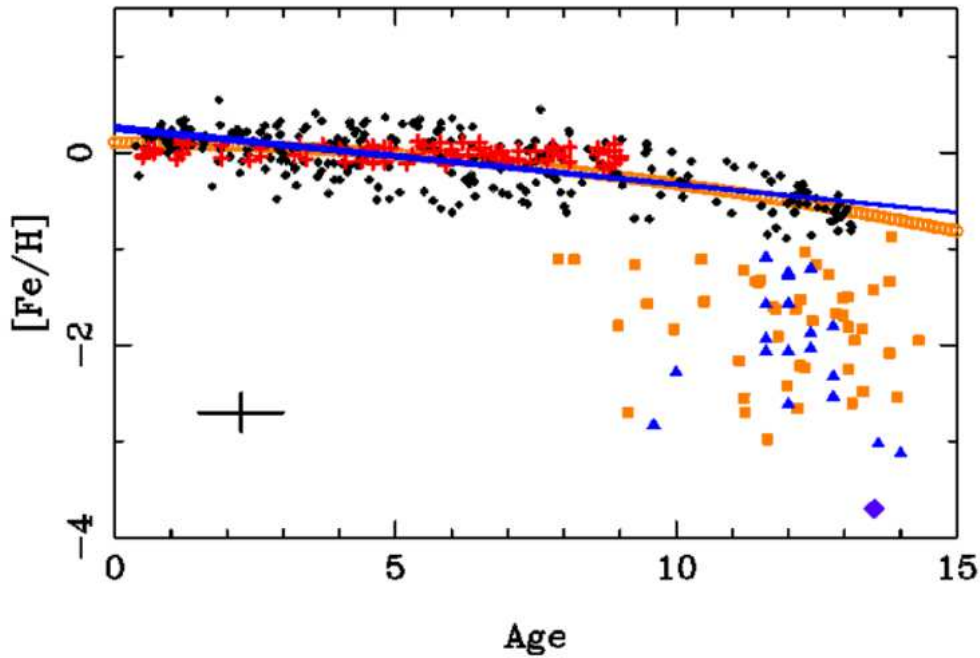}
 \centering
 \caption{[Fe/H] vs. age (Gyr) from DM19 (black points), BD18 (red plus signs), 
    RO14 (blue triangles; $[Y/Mg]$ ages, see text), and GIR21 (orange squares). 
    The large dark diamond
    is for 2MASS J18082002-5104378, the oldest known thin-disk star.  We
    use its age, 13.5 Gyr \citep{sch18} as a proxy for the age of the Galaxy.
    The horizontal bar indicates typical age uncertainties.  The vertical bar
    shows is a liberal estimate ($\pm 0.2$ dex) of the largest errors of the samples (See 
    text for further details.)  The blue and orange lines are respectively an
    unconstrained least squares fit to the BD18 stars, while the wider orange
    line is a quadratic fit, constrained to have $[Fe/H] = 0$ when age is 4.6 Gyr.
    See CY22 for further details.
    }
 \label{fig:Fig1b}
\end{figure}
This is an archival study based on four surveys of FGK dwarf and subgiant stars by 
\citet[henceforth, BD18]{bd18},
\citet[henceforth, DM19]{dm19},
\citet[henceforth, GIR21]{gir21}, and
\citet[henceforth, RO14]{ro14}.
 We examine their metallicity with age, their galactic positions and motions, and compare their relative kinematics within the Galaxy, in the following sections.  Finally, we contrast our four sets of stars with results from ultra metal-poor stars.
 
\section{The data sets}
We focus on $[Fe/H]$\footnote{We use the notation
$[Y/X]$ = $\log(Y/X)_{\rm star}-\log(Y/X)_{\rm Sun}$} 
in our consideration of metallicity vs. age.
Typical values of $[Fe/H]$ decrease, and stellar age increase from the first to the 
last sample studied. 

\subsection{BD18}
BD18, and \citet{spin18} obtained precision differential abundances for
79 solar (or near solar) twins, all located within 100 pc of the Sun.  Ages, from Spina,
et al. were considered uncertain by \citet[henceforth, CY22]{cy22} roughly 1.4 Gyr, 
from a comparison of
overlapping stars in BD18 and \citet{nis20}.  The oldest stars in this set are 
$\le 9$ Gyr, and the lowest $[Fe/H]$ is -0.15 dex\footnote{The notation ``dex'' means
$10^{\rm dex}$}.  Within this age spread, there is scatter in $[Fe/H]$, but little 
indication of of an overall trend.
Abundance error estimates are very small ($\pm 0.02$ dex or less).
It was also
shown by CY22 that {\em mean} elemental abundances in this sample are well defined as a function of age, and that individual deviations from the mean abundances for 30 elements may be shown graphically (see Fig.2 of CY22).  We also note the useful correlation of 
[Eu/Ba] with age among these stars (\citet{CoSt22}). These abundance deviations may be attributed to pollution by Galactic streams and/or diffusion into and out of the solar neighborhood \citep[see][]{hel20, bon21}. 

\subsection{DM19}
The major work by DM19 involved more than 1,000 FGK stars.  We examine a subset of 
the oldest (277) sunlike dwarfs with age errors $\le 1.5$ Gyr.  Most (79\%) of the stars are
thin disk, but {\ a few} thick disk and halo objects occur among these stars. 
Uncertainties in $[Fe/H]$ were
estimated  by CY22 to be $\approx 0.11$ dex from a comparison of stars in common 
with BD18.  Importantly, the age range of the DM19 stars extends to slightly more
than 13 Gyr, and the expected average decrease of $[Fe/H]$ with age becomes evident. 
Fitting parameters to the DM19 data also reflect this age-dependence (see Fig.1 
and remarks below). 

\subsection{Titans}
GIR21 provide high quality stellar parameters and abundances 
and ages for 48 dwarfs and subgiants chosen
from \citet{cas10} and the GAIA Benchmark stars \citep{hei15}. 
Age uncertainties average 1.2 Gyr.  We estimate errors in $[Fe/H]$ to be 0.02 dex
from abundance differences in Fe \I\, and Fe \II\, lines.  The 'Titans' set is the
most accurate in $[Fe/H]$ of our four samples.  GIR21 note that Titans were the offspring of Gaia in Greek mythology.

\subsection{RO14}
We examine a subset of 19 dwarfs and subgiants from RO14's analysis of 313
metal-poor stars from the survey of \citet{bps92}.
We assess typical $[Fe/H]$ 
uncertainties in this pre-GAIA work to be of the order of $\pm 0.2$ dex for $[Fe/H]$,
based on several estimates in RO14, as well as a comparison with the 
one  star, HD 193901, in common with the Titans.  
Individual stellar ages were not determined in RO 14.  We have made crude estimates
by requiring the mean ages and their variances 
to be equal to those of the Titans.  We used the time-sensitive \citep{ber22} $[Y/Mg]$
rato, with constants appropriate to a match with the Titans:  
${\rm age} = -4.0*[Y/Mg]+ 10.0$. 
The standard deviation of the
plotted RO14 points is 1.03 Gyr, and is surely uncertain by -2 or 
+ 1.5 Gyr.
\begin{figure} 
 \includegraphics[width=14cm,height=10cm,keepaspectratio]{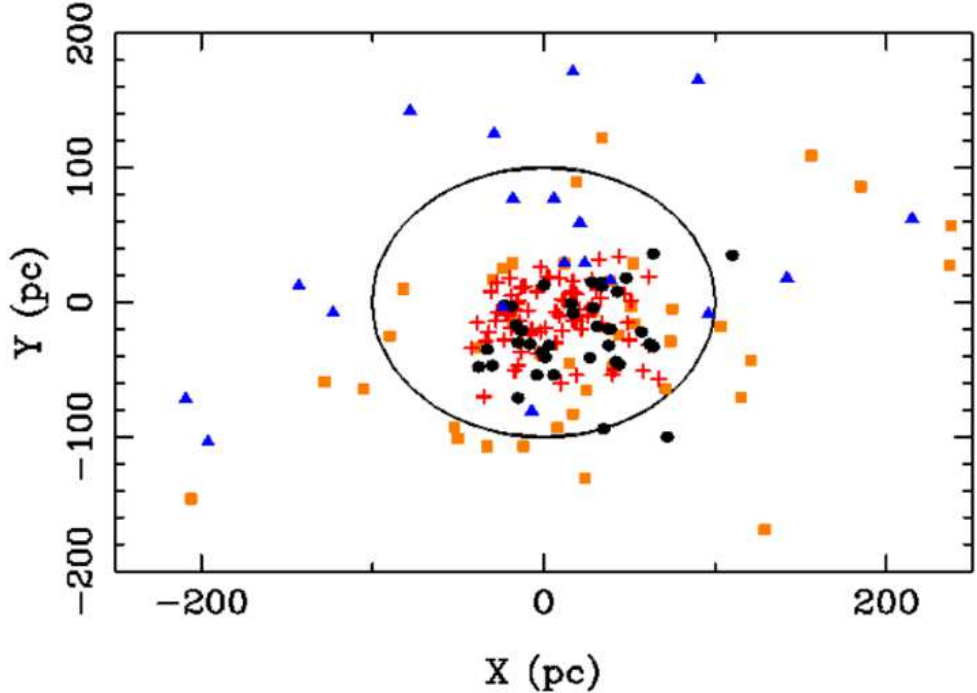}
 \centering
 \caption{Positions in the Galactic plane for four stellar data sets.
    The positive X-axis is in the direction of the Galactic center
    while the positive Y-axis is the direction of Galactic rotation.  The circle marks 100 parsecs distance.  
    Color coding and symbols
    are as in Fig.~\ref{fig:Fig1b} -- DM19 (black points), BD18 (red plus signs), 
    RO14 (blue triangles), and GIR21 (orange squares).}
 \label{fig:Fig2}
\end{figure}

\section{More distant metal-poor stars}
\citet[henceforth, PL23]{pl23}
is a study of a selection of stars from the metal-poor survey of
\citet[henceforth, BAR05]{bar05}.  The latter study included some 
250 stars with $[Fe/H] < -1.5$ as
estimated from medium resolution spectra.  
In an earlier version, PL22,  selected 28 stars for which 
age estimates with uncertainties under 1 Gyr could be obtained (see their Fig.1 and Table 10).
Uncertainties in $[Fe/H]$ were estimated as under 0.2 dex by \citet{bar05}.  This
uncertainty is small relative to the spread in the values of the PL23 stars, but 
is consonant with the vertical error bar in our Fig.1.

Although the distribution of their 28 stars would fit in Fig.1, 
the kinematics of the larger set do not.
The spread in both position and velocity space of the 252 stars of 
\citet{bar05} dwarf the spread of $Y$ vs $X$ and $U$ vs. $V$ among our four sets
(see Figs. 2 and 3).  The
distribution in space of the 252 BAR05 stars is shown in Fig. 1 of PL22.  
The distances
scatter to some 10 kpc.  In velocity space, the spread in V is beyond -1000 km s$^{-1}$, well beyond the scatter of the DM19 or RO14 points.  Further
consideration is left for future work.

\begin{figure}
 \includegraphics[width=14cm,height=10cm,keepaspectratio]{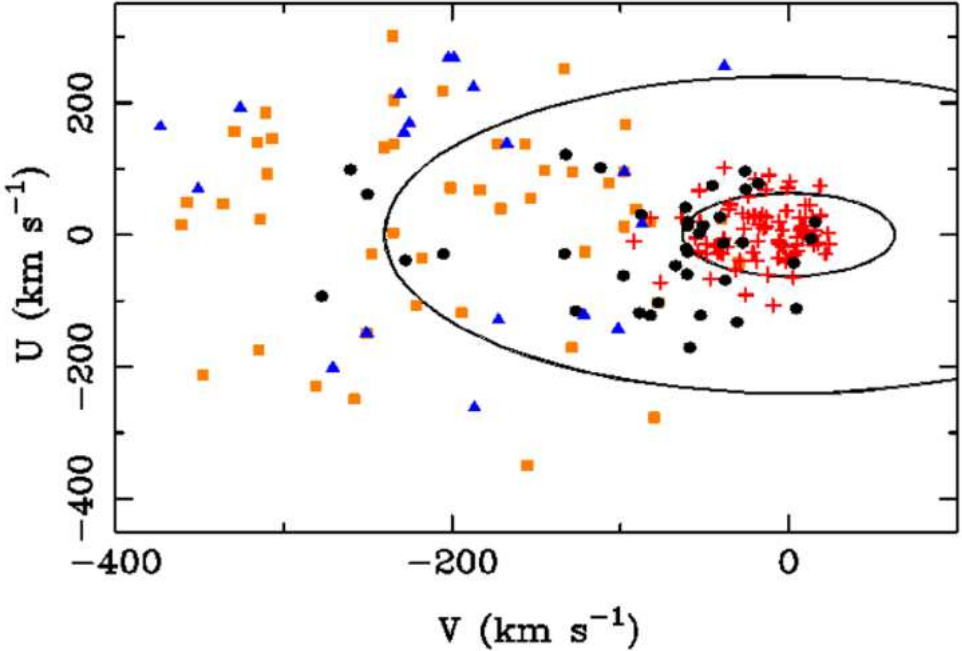}
 \centering
     \caption{Four sets of stars in velocity space.  The inner (distorted) circle
    has a radius of 63 km s$^{-1}$, the outer circle's radius is 240 km s$^{-1}$.
    Positive $U$ is in the direction of the Galactic center, while positive $V$
    is in the direction of Galactic rotation.
    Color coding and symbols
    are as in Fig.1 -- DM19 (black points), BD18 (red plus signs), RO14 (blue triangles), and GIR21 (orange squares).}
    \label{fig:Fig3}
\end{figure}  

\section{Ages}
The overall structure of the points in Fig.1 is a downward
sloping scatter of $[Fe/H]$ with age.  The scatter increases with age.  
It is uncertain whether the abrupt increase near ages of 7-8 Gyr is real
or due to systematic effects, some of which are:
\\

\begin{itemize}
\item The BD18 and DM19 stars were chosen to be similar to the sun.  Their 
vertical scatter is small but indicates real abundance variations due to pollution from Galactic streams or diffusion of stars into and out of the solar neighborhood.
\item The nineteen star RO14 sample was chosen from surveys for metal-poor stars.  A vertical separation of these stars from the BD18 and DM19 objects is therefore expected--mildly metal-weak objects were not selected for study.
\item The Titans (GIR21) were selected from the Gaia benchmark stars \citet{hei15},
but they do include benchmark stars with essentially solar $[Fe/H]$, much like Procyon.  
\end{itemize}                                                        

\section{KINEMATICS \& DYNAMICS}
The relationships among the four sets of stars may be further illustrated by
their kinematics.  Fig.2 shows the distribution of the stars
in the Galactic plane. 

The BD18 (crosses) and nearly all of the DM19 points are within 100 pc. 
None of the stars are as much as 1 kpc from the Sun.  The spread of points is smallest for the youngest (BD18) population, and widens for the oldest two.  The older DM19 stars are intermediate, as expected.

The population associations are even clearer in Fig.\ref{fig:Fig3}.  
Corrections to the local standard of rest (8.5, 13.58, and 6.49 km s$^{-1}$)
were taken from PyAstronomy\footnote{Code site: https://github.com/sczesla/PyAstronomy  }, 
whose code was also used to check 
our independent routine for transformation from equatorial to Galactic coordinates. 

The three DM19
stars outside the outer (240 km s$^{-1}$ ring are: HD 113679, HD 121004, and 
HD 148816.  They were classified halo, halo, and thick disk respectively
by DM19.  However all three are within 140 pc of the Sun.

Mean abundances of the majority of stars, which are thin-disk objects, are well 
defined as a function of age from 
0 to $\approx 10$ Gyr.  Departures from those means abundances have been illustrated 
by CY22 for the 30 elements in the
BD18 sample (see CY22, their Figure 2).\footnote{cf.  
https://zenodo.org/record/6077735.ZBtNBNLMLlg \\ -- see ObsandPredPlots .pptx (or .pdf) } . \\

These departures from the mean, for a given age,  may be attributed to the influence of 
streams and perhaps diffusion into and out of the solar neighborhood.  The transition from the solar neighborhood to the thin and thick disk, including a few halo stars, is illustrated using the samples of BD18, DM19, RO14, andGIR21.

\subsection{Phase space}
\begin{figure}
 \includegraphics[width=14cm,height=10cm,keepaspectratio]{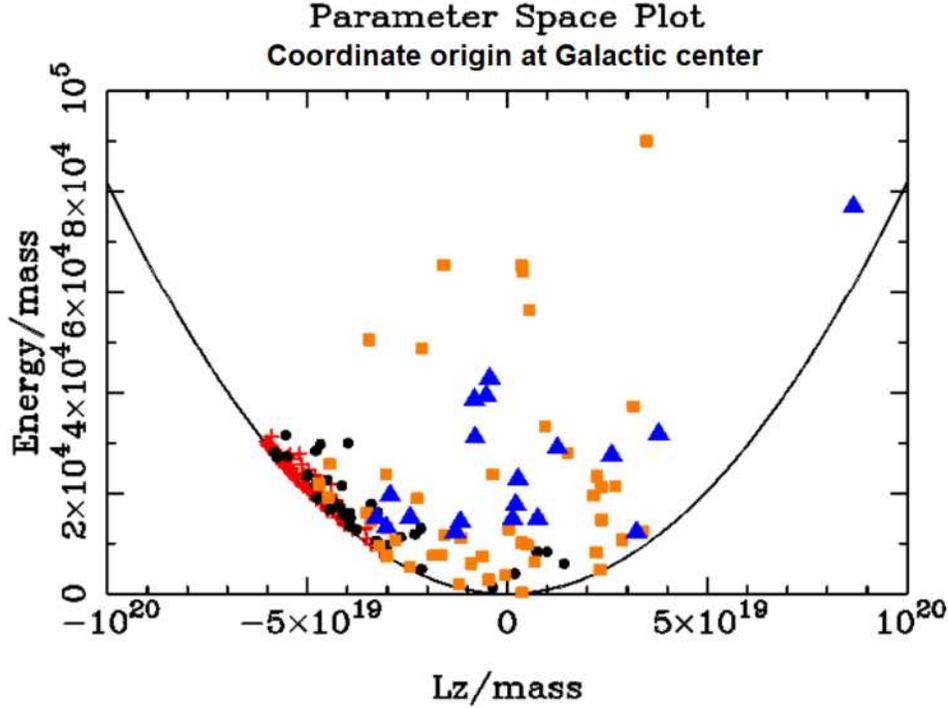}
 \centering
     \caption{Four sets of stars in parameter space. The parabola is the
    locus of points having circular velocity at the LSR.  
    See text for discussion.
    Color coding and symbols
    are as in Fig.~\ref{fig:Fig1b} -- DM19 (black points), BD18 (red plus signs), RO14 (blue triangles), and GIR21 (orange squares).}
    \label{fig:Fig4}
\end{figure}

Plots of kinetic energy,  En, versus angular momentum, Lz, highlight different aspects of Galactic kinematics \citep{bon21}.  For a central potential Lz is one of the constant integrals of motion.  The kinetic energy, of course, differs from the constant total
energy by the potential at the position of the individual stars.  Its use here avoids 
the problem of estimating that potential energy, and enables a simple, parabolic relation
between energy and Lz.  
While Figs. 2 and 3 have origins at the sun and the local standard of rest (LSR),
the coordinate origin for Fig. 4 is the Galactic center. For stars in the solar neighborhood,
\begin{equation}
En = \frac{1}{2R_c}\times Lz^2,
\label{eq:parabola}
\end{equation}
where $R_c$ is the distance to the Galactic center. 

In order to transform the coordinates of Fig.~\ref{fig:Fig1b} and Fig.~\ref{fig:Fig2}
to the Galactic center, it is only necessary to substitute $X = X-R_c$, for the 
X-coordinate, and $V = V + V_c$, where $V_c$ is the circular velocity (220 km s$^{-1}$ 
at the Local Center of Rest (LSR).  We use 8 kpc for {\bf $R_c$} \citep{grv18}, and 229 km/sec
for the circular velocity at that distance \citep{eil19}. 

\begin{figure}
 \includegraphics[width=14cm,height=10cm,keepaspectratio]{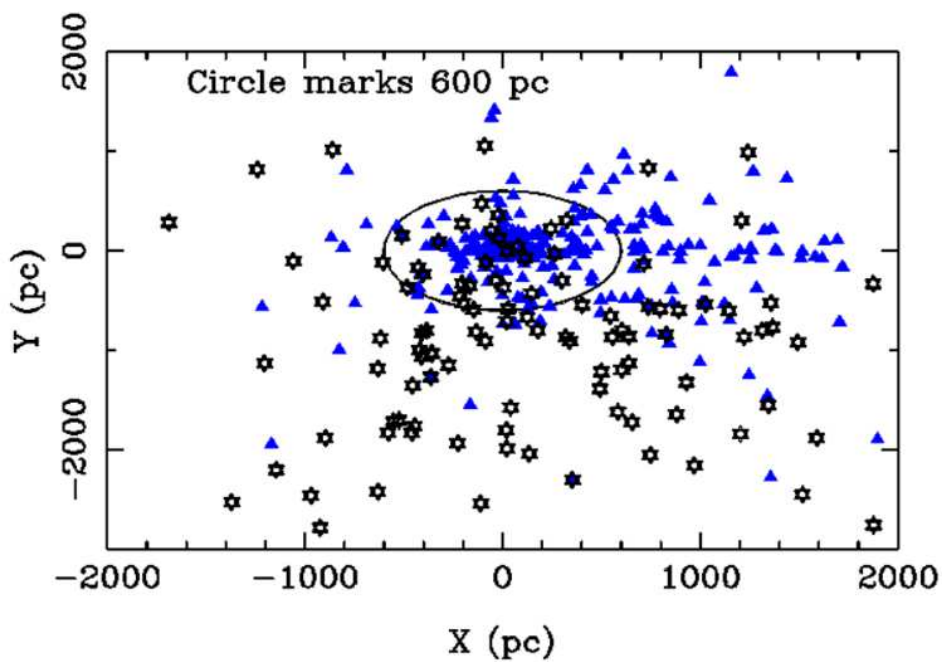}
 \centering
     \caption{Two sets of ultra metal-poor stars in the Galactic plane (Y vs X). Blue triangles
     are from RO14, including giants as well as dwarfs.  Black stars are from the Hamburg 
     survey \citep{bar05}.  The stars of these surveys reach distances significantly greater than
     the stars discussed earlier. 
    \label{fig:Fig5}}
\end{figure}  
\begin{figure}
 \includegraphics[width=14cm,height=10cm,keepaspectratio]{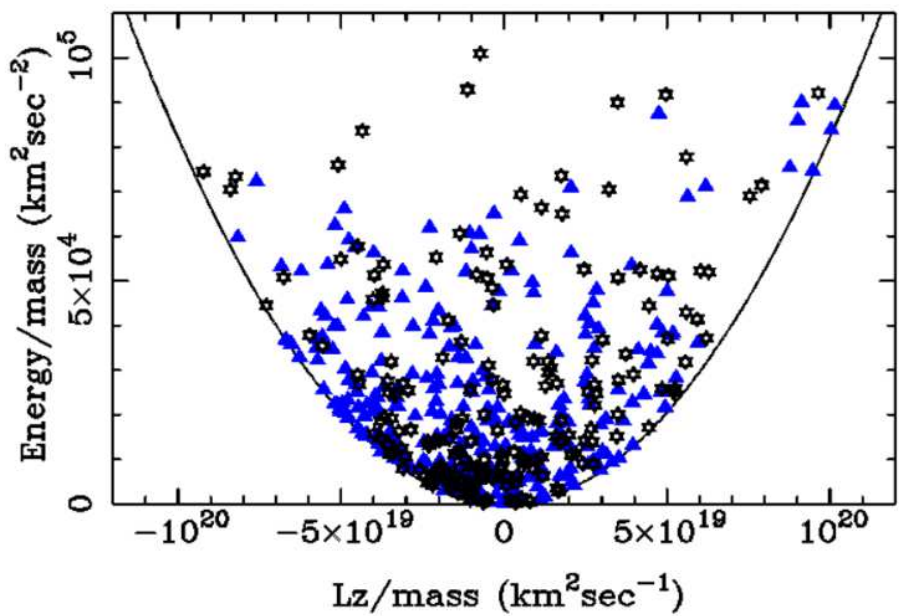}
 \centering
     \caption{En vs Lz distributions of of the RO14 and \citet{bar05} stars fill
     the parabola of Eq.~\ref{eq:parabola}. 
    \label{fig:Fig6}}
\end{figure}

The BD18 points in Figure \ref{fig:Fig4} (bright red plus
signs) cling closely to the left branch of that parabola, as expected for objects with
velocities close to the solar velocity.
The right-hand branch of the parabola is mostly unoccupied, as would be expected since
it indicates {\em retrograde} velocities.

Many of the DM points (black circles) also fall near the the BD18 clump, but some diverge
at both higher and lower energies.  Points within the parabola have, from left to right,
increasingly retrograde orbits.  This is the case for the RO14 and GIR21 stars.

Many of the GIR21 stars are members of well-known streams.  
Possible groups including RO14 stars are not noticable in this plot.

\section{Ultra metal-poor stars}
In order to put the preceeding sets of stars in perspective, we show two plots
with the distributions of ultra metal-poor stars in the Galactic plane 
and in phase space.  In these plots, we use stars of all types, giants as well as dwarfs,  
from RO14 and the Hamburg survey \citep{bar05}.  
The RO14 stars plotted here include the full 
complement of stars, not just dwarfs and subgiants as in our earlier plot, Fig.2.  
Fig. 5 shows that the positions of these stars range from 
the solar neighborhood to distances of several kiloparsecs.  
Similarly, the distributions in phase space, Fig.  \ref{fig:Fig6}, 
virtually fill the phase-space parabola, having nearly as many retrograde as
prograde orbits.  Note the accumulation of points toward the base of the 
parabola, indicating stars with both low kinetic energy and z-component
angular momentum.  
\\

\section{Summary\label{sec:sum}}
We conclude that the sample stars discussed vary systematically in $[Fe/H]$ versus age, as
well as in position and velocity space.  The plot in phase space shows a systematic
shift from nearly circular orbits in the solar neighborhood, to sub-solar and retrograde
orbits.
\section{Data availability}
The data underlying this article are available in the articles referenced and new
data may be found at in Zenodo at \url{ https://zenodo.org/record/7215968.ZBtNrNLMLlg } . 

\begin{acknowledgements}
CRC thanks E. Delgado Mena for pointing out that her (DM19) $[Fe/H]$ vs. age
decreases for the highest ages, as is expected from Galactic chemical
evolution.  We thank L. Spina for advice on the use of $[Y/Mg]$ as an age indicator.
We also thank Ian Roederer for comments and advice.
We are grateful for the advice of M. Valluti and M Chiba for help with the phase
space section.
This work made use of PyAstronomy 
(\url{https://github.com/sczesla/PyAstronomy} ).  
RES thanks the Simons Foundation for support of ArXiv.org. 
This work presents results from the European Space Agency (ESA) space mission Gaia.
The Gaia archive website is \url{https://archives.esac.esa.int/gaia }.
We also acknowledge use of the SIMBAD database \citep{wen00} operated at
CDS, Strasbourg, France. 

\end{acknowledgements}

\end{document}